\documentstyle[prd,aps,floats]{revtex}
\begin{document}

\draft

%%%%%%%%%%%%%%%%%%%%%%%%%%%%%%%%%%%%%%%%%%%%%%%%%%%%%%%%%%%%%%%%%%%%%%
%
%  Uncomment following four lines and one below for 2 column format
%  and figure insertions.
%
\input epsf
\renewcommand{\topfraction}{0.8}

%  This is the first line to be uncommented for 2 column format
%@twocolumnfalse\endcsname
%%%%%%%%%%%%%%%%%%%%%%%%%%%%%%%%%%%%%%%%%%%%%%%%%%%%%%%%%%%%%%%%%%%%%%

\preprint{0000-00}

\title{Waveform Propagation in Black Hole Spacetimes: \break
Evaluating the Quality of Numerical Solutions}

\author{L.~Rezzolla~$^{1,4}$, A.~M.~Abrahams~$^{4,6}$,
	T.~W.~Baumgarte~$^{1}$, G.~B.~Cook~$^{2}$, 
	M.~A.~Scheel~$^{2}$, S.~L.~Shapiro~$^{1,3,4}$ and 
	S.~A.~Teukolsky~$^{2,5}$.}

\address{$^1$Department of Physics, University of Illinois at
	Urbana-Champaign, Urbana, Illinois 61801}
\address{$^2$Center for Radiophysics and Space Research, Cornell
	University, Ithaca New York 14853}
\address{$^3$Department of Astronomy, University of Illinois at
	Urbana-Champaign, Urbana, Illinois 61801}
\address{$^4$NCSA, University of Illinois at
	Urbana-Champaign, Urbana, Illinois 61801}
\address{$^5$Departments of Physics and Astronomy, Cornell
	University, Ithaca New York 14853}
\address{$^6$Morgan Guarantee Trust, 60 Wall St., New York, New York 10260}

\maketitle
%------------------!
\begin{abstract}  %!
%------------------!

We compute the propagation and scattering of linear gravitational
waves off a Schwarzschild black hole using a numerical code which
solves a generalization of the Zerilli equation to a three dimensional
cartesian coordinate system. Since the solution to this problem is
well understood it represents a very good testbed for evaluating our
ability to perform three dimensional computations of gravitational
waves in spacetimes in which a black hole event horizon is present.

%------------------!
\end{abstract}    %!
%------------------!

\pacs{PACS numbers: 0.4.70.Bw, 04.25.Dm, 04.25.Nx, 04.30.Nk \hspace*{2mm} 
%Preprint 00000-AZ 
}
 
%  This is the other line to be uncommented for 2 column format
%\vskip2pc]

%%%%%%%%%%%%%%%%%%%%%%%%%%%%%%%%%%%%%%%%%%%%%%%%%%%%%%%%%%%%%%%%%%%%%%
\section{Introduction}
\label{introduction}
%%%%%%%%%%%%%%%%%%%%%%%%%%%%%%%%%%%%%%%%%%%%%%%%%%%%%%%%%%%%%%%%%%%%%%

	Binary black hole systems are among the most promising sources for
gravitational wave detectors currently under construction such as
LIGO, VIRGO and GEO.  Theoretical templates of the waveform emitted
during the inspiral and coalescence of such binaries are needed both
for increasing the probability of a detection and for extracting
astrophysical information from the signal~\cite{cetal93}. The
prediction of such waveforms has therefore become an important task of
numerical relativity and is the goal of the Binary Black Hole
Grand Challenge {\it Alliance} \cite{bbh}.

	In general, the solution of Einstein's field equations, a
large set of coupled nonlinear partial differential equations, is a
task of considerable difficulty.  Additional complexity is introduced
by the presence of singularities and the black hole horizon boundaries
necessitated by these singularities.  Most significantly perhaps, the
computational resource limitations, even on large parallel systems,
put severe constraints on the accuracy achievable with
three-dimensional (3D) simulations \cite{note1}. As a result, the
problem is far from being solved in general and, at best, there are
tailored solutions to specific difficulties.

	Current successful 3D numerical relativity computations
involve either Schwarzschild black holes (i.e. in the absence of
gravitational radiation) or gravitational waves on spacetimes where no
black holes are present. In our calculation we combine both components
and study the propagation of 3D linearized waves in the fixed
background of a Schwarzschild black hole. The solution to this problem
is well understood; it has been extensively investigated in the past
via perturbation theory \cite{rw57,z70a,z70b,v70a,v70b,p72,m74} and
numerous 1D calculations have also been made
\cite{cpm79,aetal92,aetal93,gpp94,pp94,ac94,ast95,ap96,getal96}.  In
particular, the gravitational waves can be treated as three
dimensional perturbations of the background metric and be expanded in
terms of tensor spherical harmonics. This reduces the problem to
solving a 1D radial equation for each component of the tensor
spherical harmonics. For odd parity perturbations this radial equation
is known as the Regge-Wheeler equation, while for even parity
perturbations it is the Zerilli equation. An alternative,
non-perturbative approach has employed the matching of a Cauchy
solution of Einstein's equations onto characteristic hypersurfaces
\cite{betal97}.

	Here, to test our ability to track gravitational radiation
numerically in a 3D black hole spacetime, we (artificially)
reintroduce the angular dependence into the Zerilli equation and
evolve the resulting equation in three dimensions. Many of the
difficulties which have beset a fully self-consistent, nonlinear
calculation are absent in this test problem. In particular, the
equations are linear and the location of the black hole is known
independently of our numerical solution. At the same time, our test
problem shares many features of the full nonlinear problem. For
example, we excise the black hole from the computational grid and
impose boundary conditions on the apparent horizon. In addition, we
use the same computational infrastructure usually implemented in
present 3D numerical relativity codes, which allows both for parallel
applications and, in principle, adaptive mesh refinement. Finally,
since we can solve the 1D Zerilli equation with almost arbitrary
resolution, we can always compare our 3D results with an essentially
``analytic'' solution.

	The point of this paper is to present this testbed problem
\cite{idea} and demonstrate its potential usefulness as a calibrater of
numerical accuracy for wave propagation in 3D black hole spacetimes.
The numerical exercises we perform are illustrative only and our
implementation is somewhat crude; also, we do not fully exploit the 
capabilities of the largest parallel machines. Our examples are
sufficient, however, to convey the basic idea so that future code
builders can already begin utilizing this tool in their 3D diagnostic
work.

	The paper is organized as follows: in Section \ref{tze} we
review the key equations. Section \ref{bc} is devoted to a discussion
of the different boundary conditions used in the computation, while in
Section \ref{niam} we give a brief description of the 3D code and of
our choices for the numerical implementation. Illustrative numerical
tests and results are presented in Section \ref{nrsogr}, where we also
discuss the relevance of appropriate inner and outer boundary
conditions.  Finally, Section \ref{conclusion} contains our
conclusions and suggestions for future extensions.

%%%%%%%%%%%%%%%%%%%%%%%%%%%%%%%%%%%%%%%%%%%%%%%%%%%%%%%%%%%%%%%%%%%%%%
\section{The Zerilli Equation}   
\label{tze}
%%%%%%%%%%%%%%%%%%%%%%%%%%%%%%%%%%%%%%%%%%%%%%%%%%%%%%%%%%%%%%%%%%%%%%

	Regge and Wheeler \cite{rw57}, examining the stability of the
Schwarzschild geometry against small nonspherical perturbations,
introduced a decomposition of the perturbations in terms of tensor
spherical harmonics. Splitting the tensor harmonics into ``even'' and
``odd'' parity classes, they found that it was possible to write the
equations of the odd-parity harmonics in terms of a single second
order differential equation describing wave motion in an effective
potential (i.e. the Regge-Wheeler equation). Zerilli \cite{z70a}
reconsidered the problem and found the corresponding equation for
even-parity tensor harmonics (i.e. the Zerilli equation). Finally,
Moncrief \cite{m74} derived versions of both even and odd-parity
equations using an elegant gauge-invariant formulation of the
perturbations.  Here we concentrate on the even-parity perturbations
and describe them in terms of the Zerilli equation,

\begin{equation}
\label{zer_t}
{\partial^2 Q_{\ell} \over \partial t^2} - 
{\partial^2 Q_{\ell} \over \partial r^2_*} + V_{\ell}(R) Q_{\ell} 
= 0 \ ,
\end{equation}
where $t$ and $R$ are the time and radial Schwarzschild coordinates
respectively and $r_*$ is the so-called ``tortoise'' coordinate,
defined as
\begin{equation}
\label{r*toRs}
r_*  =  R + 2M {\rm ln} \left({R \over 2M}  -1 \right) \ .
\end{equation}

	The Zerilli function $Q_{\ell}$ is, in typical applications,
constructed from metric perturbations and their derivatives.  The
Zerilli potential $V_{\ell}(R)$ is given by

\begin{equation}
V_{\ell}(R) = 
N^2\left\{ {1\over\Lambda^2}\left[{72 M^2\over R^5} - {12 M\over R^3}
(\ell - 1)(\ell +2)\left(1-{3M\over R} \right) \right]+ 
{\ell (\ell - 1)(\ell + 1)(\ell + 2) \over{\Lambda R^2}} \right \} \,
\end{equation}
where we have used the abbreviations
\begin{equation}
\Lambda  =  (\ell - 1)(\ell +2) + 6{M\over R} \ 
\end{equation}
and 
\begin{equation}
N^2  =  \left(1-{2M\over R}\right) \ .
\end{equation}

	In order to generalize the Zerilli equation to three
dimensions, we first replace this tortoise coordinate by the
Schwarzschild radial coordinate $R$

\begin{equation}
\label{zer_s}
{\partial^2 Q_{\ell} \over \partial t^2} - 
N^4 {\partial^2 Q_{\ell} \over \partial R^2} -
{2 M N^2 \over R} {\partial Q_{\ell} \over \partial R} 
+ V_{\ell} Q_{\ell} = 0.
\end{equation}

%	In terms of the tortoise coordinate, the event horizon is at
%negative infinity (thus excluded from the computational domain) and
%the characteristic speeds are always unity. The coordinate stretching
%produced by (\ref{r*toRs}) counteracts the vanishing of the $g_{00}$
%component of the underlying Schwarzschild metric. This allows for a
%straightforward implementation of finite difference schemes both in
%the interior and close to the horizon (i.e. at sufficiently large
%negative value of $r_*$). Transforming to a Schwarzschild coordinate
%introduces $N^4$ and $N^2$ factors in front of the spatial
%derivatives, which vanish at the event horizon. These factors reflect
%the ``freezing'' of the constant coordinate-time slices and cause the
%characteristics to ``pile up'' in front of the event
%horizon~\cite{fn1}.  This results in problems for a numerical
%implementation since it leads to the formation of arbitrarily small
%features in the solution which will ultimately be smaller than any
%grid resolution. Straightforward ingoing wave boundary conditions will
%not be adequate in these coordinates unless a sufficient grid
%refinement is achieved in the vicinity of the horizon.
%	Unfortunately, it is not obvious how to generalize a
%tortoise-like coordinate to three dimensions and accordingly the 
%apparent horizon cannot be removed from the computational grid. In the
%following we will therefore focus on Schwarzschild coordinates.

	In order to generalize equation~(\ref{zer_s}) to a 3D
Schwarzschild polar coordinate system $(t,R,\theta,\phi)$, we define a
new Zerilli function $Q=Q(t,R,\theta,\phi)$ to be the product of two
separable functions

\begin{equation}
\label{def_Q}
Q \equiv Q_{\ell}(t, R){\cal A}( \theta, \phi) \ ,
\end{equation}
where $Q_{\ell}(t, R)$ is a solution of equation (\ref{zer_s}) and
${\cal A}(\theta, \phi)$ contains the (``artificial'') angular
dependence expressed in terms of scalar spherical harmonics. We demand
that the angular part in the Zerilli function is an eigenfunction of
the perpendicular differential operator $\nabla^2_{\perp}$, i.e.

\begin{equation}
\nabla^2_{\perp} {\cal A} \equiv
\left[ {1 \over {R^2 {\rm sin} \theta}} {\partial \over \partial \theta}
\left( {\rm sin} \theta {\partial \over \partial \theta} \right) +
{1 \over {R^2 {\rm sin}^2 \theta}}{\partial \over \partial^2 \phi} 
\right ] {\cal A} = - {\ell(\ell+1) \over {R^2}} {\cal A} \ .
\end{equation}
With the transformation
\begin{equation}
\label{cart_s}
x = R~{\rm sin}\theta~{\rm cos} \phi \ , \hskip 1.0truecm
y = R~{\rm sin}\theta~{\rm sin} \phi \ , \hskip 1.0truecm
z = R~{\rm cos}\theta \ 
\end{equation}
and
\begin{equation}
{\partial \over \partial R} = {x^i \over R} 
{\partial \over \partial x^i} \ ,
\end{equation}
we then write equation~(\ref{zer_s}) in terms of 3D
Cartesian coordinates
\begin{equation}
\label{zer_3s}
{\Box}_{_S} Q \equiv \left [ {\partial^2 \over{\partial t^2}} - 
N^4 \left({\partial^2 \over{\partial x^2}} + 
	  {\partial^2 \over{\partial y^2}} +
	  {\partial^2 \over{\partial z^2}} \right) \right ] Q 
= - {2 N^2 \over R^2}\left( 1 - {3M \over R} \right) 
x^i {\partial Q\over {\partial x^i }} - 
V_{\ell,_{S}} Q \ ,
\end{equation}
where
\begin{equation}
V_{\ell,_{S}}  =  V_{\ell} - { {\ell}(\ell+1) \over R^2} N^4 \ .
\end{equation}

	In order to take advantage of evolution schemes designed
for first-order hyperbolic systems, we 
rewrite our system using the following derived quantities:
\begin{eqnarray}
\hskip 2.0truecm Q_0 & \equiv & {\partial Q \over {\partial t }}\ , \\
\hskip 2.0truecm Q_i & \equiv & {\partial Q \over {\partial x^i}} 
\hskip 2.0truecm i=1,2,3 \ . 
\end{eqnarray}
Equation~(\ref{zer_3s}) is then equivalent to the first order system
\begin{eqnarray}
\label{first_order}
\left\{
\begin{array}{rcl}
\displaystyle{ {\partial Q_0 \over {\partial t }} }
+ c_1 \sum_{i=1}^{3} \displaystyle{ {\partial Q_i \over {\partial x^i
}} } & = & c_2 x^i Q_i + c_3 Q \ ,\\ \\ 
\displaystyle{ {\partial Q_i \over {\partial t }} } -
\displaystyle{ {\partial Q_0 \over {\partial x^i }} } & = & 0 \ , 
\hskip 2.0truecm  i=1,2,3  \ , \\ \\ 
\displaystyle{ {\partial Q \over {\partial t }} }& = & Q_0 \ \ , \\
\end{array}
\right . 
\end{eqnarray}
where we have used
\begin{equation}
c_1 = - \left(1-{2M\over R}\right)^2 \ ,  \hskip 1.5truecm 
c_2 = - {2 \over R^2} \left(1-{2M\over R}\right) 
        \left( 1 - {3M \over R} \right) \ ,  \hskip 1.5truecm 
c_3 = - V_{\ell} + { \ell (\ell+1) \over R^2} c_1 \ .   
\end{equation}

%%%%%%%%%%%%%%%%%%%%%%%%%%%%%%%%%%%%%%%%%%%%%%%%%%%%%%%%%%%%%%%%%%%%%%
\section{Boundary Conditions}
\label{bc}
%%%%%%%%%%%%%%%%%%%%%%%%%%%%%%%%%%%%%%%%%%%%%%%%%%%%%%%%%%%%%%%%%%%%%%

	We require three different kinds of boundary conditions. The first
ones result from restricting the computational domain to one octant,
the second ones are the ``outer'' boundary conditions specified at
large distance and the third ones are the ``inner'' boundary conditions
on the apparent horizon. We will describe several different choices
for the inner and outer boundary conditions and compare results in
section~\ref{nrsogr}.

\subsection{Octant Symmetry Boundary Conditions}
\label{osbcs}

	Octant symmetry permits us to restrict our analysis to a
cubical grid in which all coordinate values are positive (this is
what we refer to as an ``octant''). As a result, boundary conditions
have to be imposed on the octant-symmetry planes and the specific
conditions on the functions $Q_0, Q_x, Q_y$ and $Q_z$ can be derived
from the symmetry of each function on each plane. For even $\ell$,
$Q$ is symmetric across each plane and hence $Q_0$ and derivatives
tangential to the plane are symmetric as well, while the derivative
perpendicular to the plane is antisymmetric.

\subsection{Outer Boundary Conditions}
\label{obcs}

At the outer boundaries we impose outgoing wave Sommerfeld conditions.
In this approximation, we assume that the functions are of the form
$\Phi = f (t-R, \theta, \phi)/R$, or equivalently,
\begin{equation}
\label{somm_sph}
{\partial \Phi \over {\partial t}} + {1 \over R}
{\partial \over {\partial R}} (R \Phi) = 0 \ .
\end{equation}
Transforming equation (\ref{somm_sph}) into a 3D Cartesian coordinate
system yields
\begin{equation}
\label{somm_cart}
{\partial \Phi \over {\partial t}} + {1 \over R}
%\sum_{i=1}^3 
x^i {\partial \Phi \over {\partial x^i}} + 
{\Phi \over R} = 0 
\end{equation}
and requires, for each grid point on the external faces of the cubical
grid, evaluating derivatives perpendicular as well as tangential to
the surface. However, if $f \sim f (t-R)$, then on most parts of
the outer surface the tangential derivatives are fairly small. It is
therefore adequate to neglect these and replace~(\ref{somm_cart}) with
the simpler expression~\cite{mc96b}

\begin{equation}
\label{somm_cart_rad}
{\partial \Phi \over {\partial t}} + 
{x^i \over R} {\partial \Phi \over {\partial x^i}} + 
{\Phi \over R} = 0 \ ,   \hskip 2.0 truecm  \ \ 
i=1,2 \mbox{~or~} 3,~~~\mbox{(no summation)}  
\end{equation}
(see Appendix \ref{aobcs} for details on the finite difference form).

	One concern which might be related with this prescription for
the outer boundary conditions is that, strictly speaking, it is going
to be satisfied only by those grid points on the outer faces which
happen to be aligned with the radial direction of propagation of the
wave-like quantities. All of the other grid points (e.g. those on the
edges and corners of the cubical grid) will not satisfy expressions
(\ref{somm_cart_rad}) identically and a certain amount of reflection
will occur.  While there are several different ways of handling these
additional complications [e.g. use of a spherical outer boundary
embedded within the cubical grid, or an explicit computation of the
radial derivative in expression (\ref{somm_sph}) through
interpolations \cite{cs97a}], experience has shown that, provided that
the outer boundary is placed at a sufficiently large distance, the
amount of reflection produced is usually very small and conditions
(\ref{somm_cart_rad}) are sufficient to provide a stable outer
boundary.

\subsection{Inner Boundary Conditions}
\label{ibc}

%	From a mathematical point of view, no boundary condition can
%be imposed at the apparent horizon. This is, because the past domain
%of dependence of any point on the apparent horizon does not include
%any point inside the horizon. Stated differently, there are no
%outgoing characteristics on the horizon along which information could
%enter our computational domain and the solution is therefore uniquely
%determined in terms of the solution outside the horizon.
%However, in a numerical implementation, the innermost grid point still
%has to be treated differently since it does not have all the
%neighbouring grid points necessary for a centered updating scheme as it
%is used in the interior.

	At the inner boundary we use a horizon excising method
which has been discussed by a number of different authors
\cite{ss92,bms95,sst95,anetal95,ay95,mc96a,setal97} in conjunction
with the implementation of apparent horizon boundary conditions.  In
general, (e.g. as in the case of moving black holes), such a region is
not known a priori and its location has to be computed at each time
step with ``apparent horizon finder'' routines~\cite{betal96,t96,getal97}.
This is not the case for the present static Schwarzschild background
and we need to excise the region inside the event horizon only once
during our time integration. Given a masked out region of spacetime,
the simplest inner boundary conditions involve using suitable finite
difference stencils and higher order extrapolation methods that fill,
on each spacelike hypersurface, the gridpoints just inside the masked
region and adjacent to the event horizon. These values can then be
used in a centered evolution scheme to update the gridpoints just
outside the horizon. We have implemented such boundary conditions
using a fourth order extrapolation scheme.  These boundary conditions
do not violate causality since no information is extracted from within
the event horizon and, when stable, provide boundary values that are
mathematically correct. This prescription is simple to implement, does
not require special assumptions on the behaviour of the variables in
the proximity of the horizon (as would, instead, a Sommerfeld
condition) and has been proven to be stable for wave propagation in
2$+$1 dimensions on a flat spacetime~\cite{ay95}.

	Alternatively, we can explicitly evaluate equation
(\ref{zer_3s}) on the horizon. Choosing $Q_i = 0$ ($i=1,2,3$)
initially, we find the following ``freezing'' conditions on the
horizon
\bigskip
\begin{eqnarray}
\label{freeze_bc}
\left\{
\begin{array}{rcl}
{\partial_t } Q_0 = 0 \ &,& \  \\ \\ 
{\partial_t } Q_i = 0 \ &,& \   i=1,2,3  \ .
\end{array}
\right . ~~~~\mbox{at R = 2M}
\end{eqnarray}
These conditions do not involve extrapolation, are very easy to
implement into the adopted Macormack evolution scheme (see
Appendix~\ref{afde}) and as we will show in Section~\ref{nrsogr},
produce more reliable results than the more general boundary condition
based on extrapolations. A different choice of time coordinate may not
necessitate such care on the horizon.

%%%%%%%%%%%%%%%%%%%%%%%%%%%%%%%%%%%%%%%%%%%%%%%%%%%%%%%%%%%%%%%%%%%%%%
\section{Numerical Implementation}
\label{niam}
%%%%%%%%%%%%%%%%%%%%%%%%%%%%%%%%%%%%%%%%%%%%%%%%%%%%%%%%%%%%%%%%%%%%%%
	
	We use different codes to solve the 1D and the 3D Zerilli
equations. Since the 1D code can be run with an essentially arbitrary
number of gridpoints and hence essentially arbitrary accuracy, we use
this solution as an ``analytic'' solution for comparisons.  Moreover,
since the 1D code is considerably simpler, we used this code to
experiment with several computational techniques before implementing
them in the 3D code.

	The 3D code implements a time evolution scheme for
equations~(\ref{first_order}) in an environment very similar to other
codes of the Alliance, in particular the so-called ``Empire'' code.
The latter evolves Einstein's equations in a hyperbolic formulation
\cite{cby95,aetal95}, so that the mathematical structure of these
equations is identical to that of (\ref{first_order}). Our code has
been implemented using the DAGH environment~\cite{pb95}, which has
been developed for the Alliance. DAGH is a Distributed Adaptive Grid
Hierarchy software package which allows for parallel applications and,
in principle, Adaptive Mesh Refinement (AMR). Our code runs in
parallel, but we have not implemented AMR here.  The code uses a 3D
cartesian cell-centered uniform grid of typically $(32)^3$, $(64)^3$,
$(128)^3$ or $(256)^3$ gridpoints. For most applications we restrict
the computation to a single octant and a typical run with $(128)^3$
gridpoints distributed over 16 processors of the Origin2000 at NCSA
would evolve up to $t = 100\;M$ in about 6 hours of CPU time.

	Both the 1D and the 3D codes use a Macormack evolution scheme.
In this algorithm a ``predictor-step'' evolves the variables, to
linear order, from time level $t$ to the subsequent time level $t +
\Delta t$ and a ``corrector-step'' uses both the old and the
predicted values to improve the integration and make it second order
accurate (see Appendix \ref{afde} for the explicit finite difference
form of the equations).  All of the spatial derivatives during the
predictor and the corrector steps are one-sided and therefore only
first order accurate in space. However, combining the predictor and
corrector steps on a uniform grid cancels the first order error terms,
so that the scheme becomes second order accurate both in space and
time. 

\begin{figure}
\centerline{\epsfysize=13cm \epsfbox{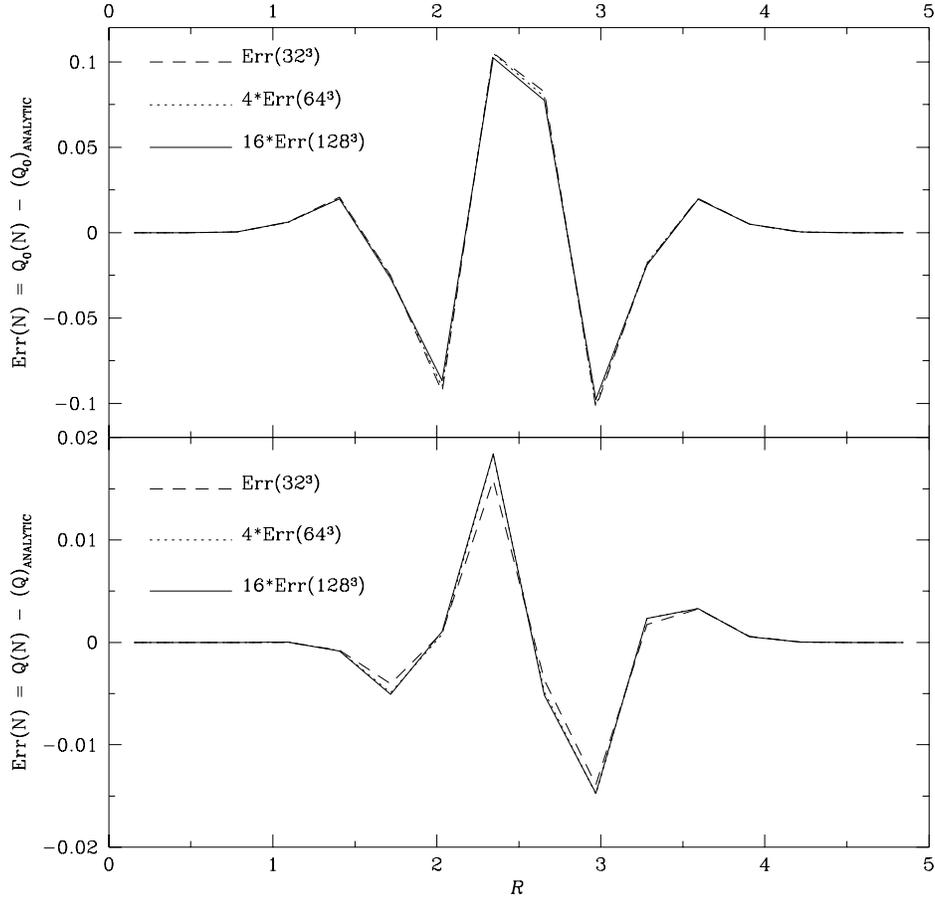}} 
\caption[fig1]{\label{fig1} 
Convergence test of the 3D code. All of the curves show the leading
(second) order truncation error for the amplitude of a spherical wave
$Q$ and of its time derivative $Q_0$. The curves refer to the solution
at time $t=3.125$ and are the result of an interpolation on a fixed
set of $(16)^3$ grid points of solutions computed using $(32)^3$,
$(64)^3$ and $(128)^3$ gridpoints respectively. In order to maintain a
consistent treatment of the interpolation error introduced, the
solutions obtained with $(16)^3$ gridpoints have not been used for
this test.}
\end{figure} 

	We verify the second order accuracy of our code in
Figure~\ref{fig1}.  For this particular test we evolved a spherical
wave on a flat background [i.e. equation~(\ref{first_order}) with
$c_1=1$ and $c_2=c_3=0$], for which the solution is known
analytically. In a second order accurate code the deviation from the
analytic solution decreases by a factor of four when the grid
resolution is doubled. As a test, we plot in Figure~\ref{fig1} the
error for three different grid resolutions and multiply the errors
with successive factors of four. The curves refer to the solution at
time $t=3.125$ which is reached after 20 time steps on the coarsest
grid and 160 time steps on the finest. Small deviations between the
curves are caused by higher order error terms. For increasing
resolution, these deviations decrease and the curves converge to a
limiting graph. This proves that our code is indeed second order
accurate.

%%%%%%%%%%%%%%%%%%%%%%%%%%%%%%%%%%%%%%%%%%%%%%%%%%%%%%%%%%%%%%%%%%%%%%
\section{Numerical Results} 
\label{nrsogr}
%%%%%%%%%%%%%%%%%%%%%%%%%%%%%%%%%%%%%%%%%%%%%%%%%%%%%%%%%%%%%%%%%%%%%%

	In this section we present numerical results of our 3D
evolutions of the Zerilli equation and in particular we compute the
scattering of gravitational radiation off a Schwarzschild black hole
\cite{v70a,v70b}.  Consider a gravitational wave packet coming from
large radius and moving towards the effective potential produced by
the black hole. As a result of the scattering process, part of the
wave packet will be transmitted across the potential and eventually
cross the event horizon and part of it will be reflected by the
potential and propagate out again to spatial infinity. As in quantum
mechanics, the transmission and reflection coefficients will depend on
the wavelength of the incoming radiation (i.e. very high frequency
modes will be almost completely transmitted while very low frequency
modes will be almost completely reflected) as well as on properties of
the scattering potential (i.e. on the mass of the black hole). The
knowledge of the rate at which the gravitational radiation is
reflected off the black hole as a function of the wavelength of the
incoming radiation would provide a distant observer with information
about the mass of the black hole.

As initial data for the Zerilli function we choose a Gaussian of width
$\sigma$ centered at radius $R_0$
\begin{equation}
Q(0,R,\theta,\phi) = {\rm exp} 
\left[- {(r_*(R)-r_*(R_0))^2 \over {\sigma^2}}\right] P_{\ell m} \,
\end{equation}
where $P_{\ell m}$ is an associated Legendre function.  For all
calculations in this section we chose $R_0 = 10 M$ and $\sigma = 1
M$. Defining this Gaussian in terms of the tortoise coordinate
guarantees that $Q$ vanishes on the horizon. We also assume time
symmetry, i.e.
\begin{equation}
{\partial Q (0,R,\theta,\phi) \over {\partial t}} 
\equiv Q_0 (0,R,\theta,\phi) = 0  
\end{equation}
and impose outer boundary conditions at $x = y = z = 20\;M$.

	As the wave packet evolves in time, it splits, with one part
of it going towards null infinity and the other reaching the horizon,
where it induces the quasi-normal ringing of the black hole. The scope
of this computation is to calculate the waveform of the scattered
gravitational radiation as observed at some distance from the black
hole and to compare it with the ``analytic'' waveform, i.e. the
solution of the 1D Zerilli equation (\ref{zer_t}).

\begin{figure}
\vbox{
\centerline{\epsfysize=12cm \epsfbox{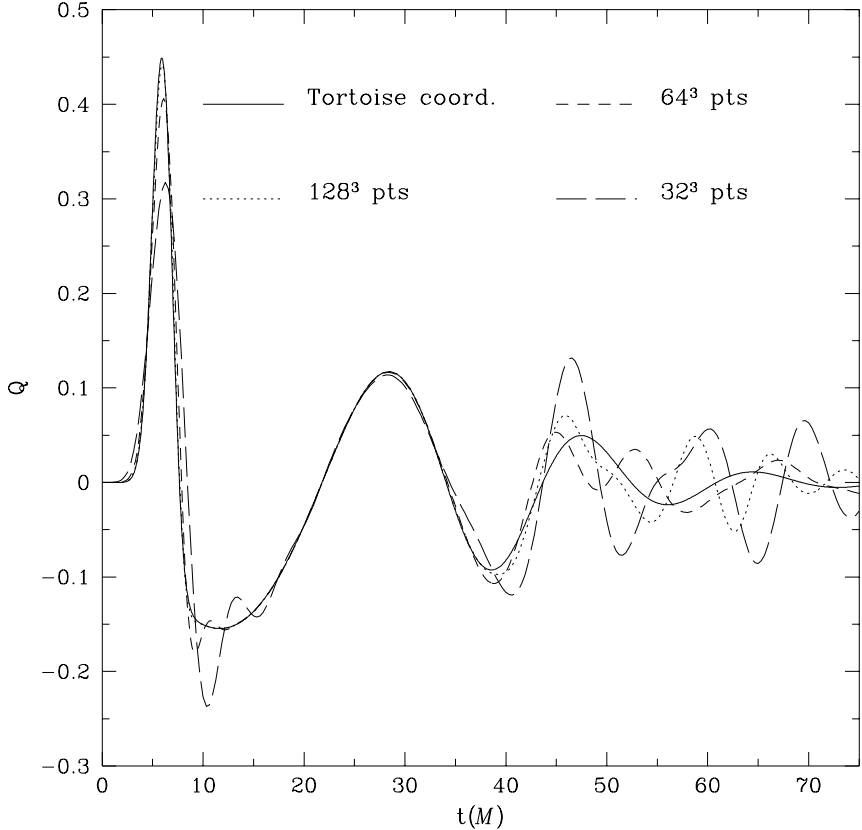}}
\caption[fig2]{\label{fig2} 
The Zerilli function $Q_2$ as a function of $t$ at radius $R=15 M$.  The
solid curve is the ``analytic solution'' and the others are 3D
solutions computed with $(32)^3$, $(64)^3$ and $(128)^3$ gridpoints.
We use one-sided differencing at the inner boundary. }}
\end{figure}

In Figure \ref{fig2} we show $Q_{\ell}$ with $\ell = 2$ as a function
of time at radius $R=15\;M$. For this calculation we used one-sided
differencing at the inner boundary. More precisely, we use a quartic
extrapolation to fill a fictitious gridpoint just inside the horizon,
so that the gridpoint just inside the horizon has all the neighbors
required for a centered updating scheme. We can then apply our normal
``interior'' evolution scheme to update this gridpoint.

Here and in the following three figures, the solid curve is the
``analytic'' solution, which we found by integrating the 1D Zerilli
equation (in tortoise coordinates) with a large number of gridpoints.
The other curves are the results of 3D computations with $(32)^3$,
$(64)^3$ and $(128)^3$ gridpoints.

The first peak and the first minimum in this curve are produced by the
part of the wave packet which moves outwards and leaves the
computational grid\cite{fn2}. The second peak and all the following
ones are the quasi-normal ringing of the black hole excited by the
infalling part of the wave packet.

	Note that for times larger than $t \approx 35 M$ our 3D
solutions do not converge to the analytic result. Higher resolution
improves the solution for early times and delays the time at which the
computed curve starts to deviate from the analytic one. This behavior
is caused by the properties of the Schwarzschild coordinates. Since
the ``lapse'' function vanishes on the horizon, infalling waves slow
down as they approach the horizon and will, in terms of the
Schwarzschild time $t$, never reach the horizon.  Stated differently,
the characteristic speeds vanish on the horizon, so that the infalling
waves pile up in front of the horizon. This causes the formation of
increasingly small features close to the horizon. These will
ultimately be smaller than any (uniform) grid resolution.  For any
constant grid resolution there is therefore a time after which
features close to the horizon can no longer be resolved. In this
region of the spacetime and in its domain of dependence, the numerical
solution will therefore be spoiled. This is an unavoidable feature of
the underlying Schwarzschild coordinates together with a uniform grid.

\begin{figure}
\centerline{\epsfysize=12cm \epsfbox{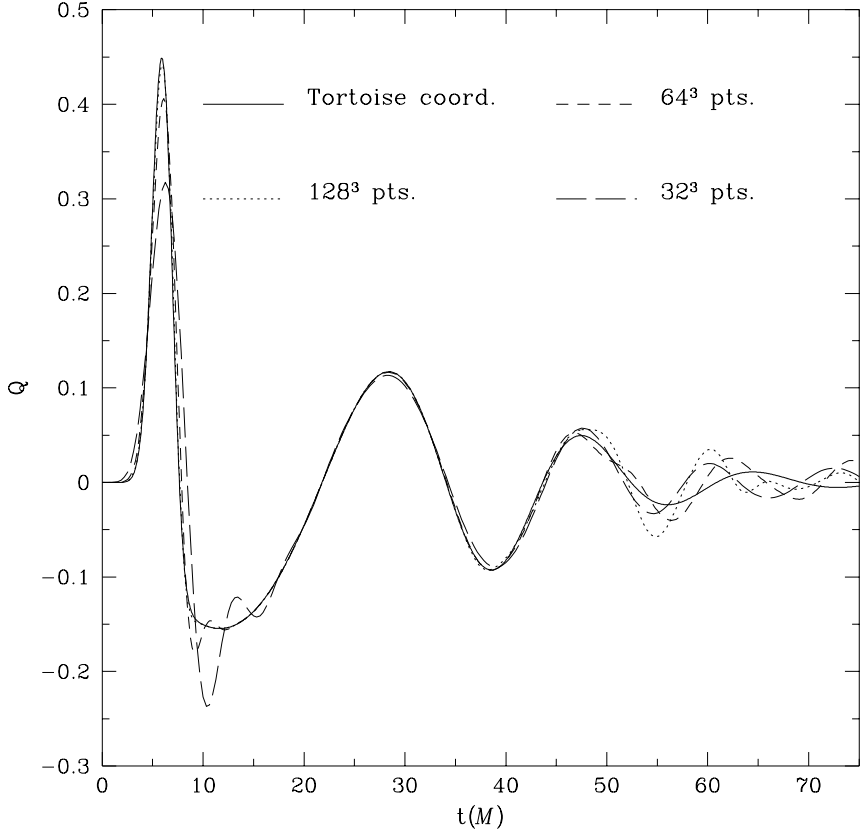}} 
\caption[fig3]{\label{fig3}
Same as Figure 2, except that we impose the ``freezing'' boundary 
conditions at the inner horizon.}
\end{figure}

	Obviously, using a high order extrapolation close to the
horizon will produce spurious results and will further decrease the
quality of the solution. In Figure~\ref{fig3} we show the results of
the same calculation, except that we imposed the ``freezing'' boundary
conditions~(\ref{freeze_bc}), which do not involve extrapolation.  As
expected, our results improve and converge to the analytic solution up
to a later time of about $t \approx 45 M$.

	After this time, the solution can only be improved by
dramatically increasing the resolution close to the horizon.  In these
coordinates, the only feasible way to increase the resolution
sufficiently, given the memory limitations of today's computers, is to
use Adaptive Mesh Refinement techniques.  We leave to future
investigation the application of AMR to this problem.

\begin{figure}
\centerline{\epsfysize=12cm \epsfbox{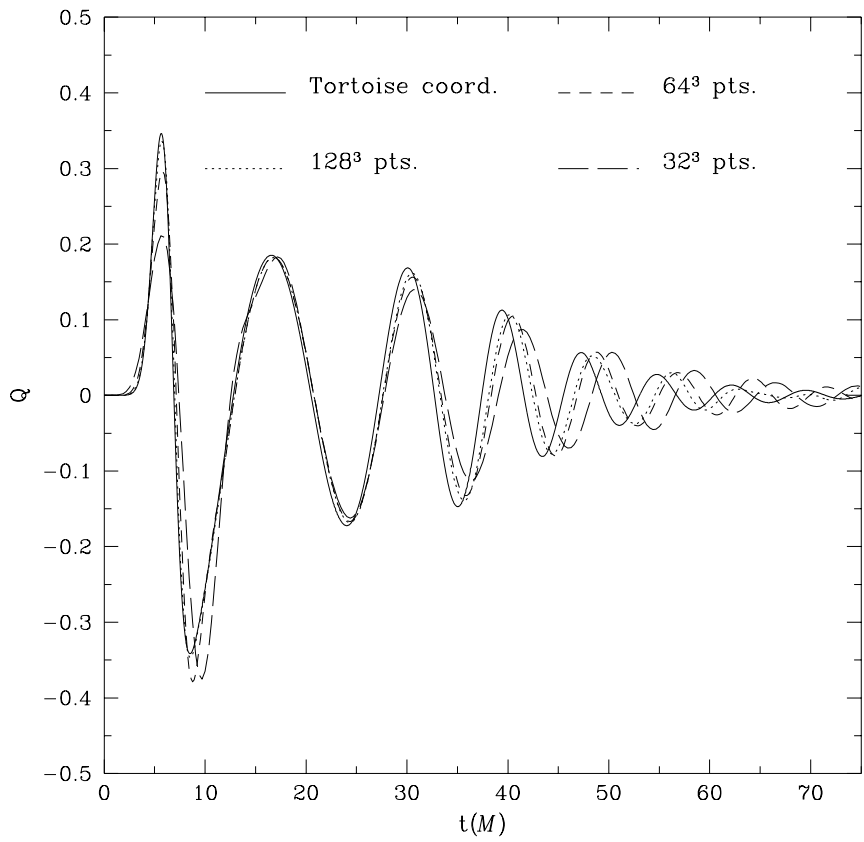}} 
\caption[fig3a]{\label{fig3a}
Same as Figure 3 for $\ell = 4$.} 
\end{figure}

	In Figure~\ref{fig3a} we show results, again using the
``freezing'' boundary conditions, for $\ell = 4$. Since these waves
have smaller structures and features to begin with, the numerical
solution becomes unreliable at an even earlier time. Note, however,
that these deviations are mostly due to a phase error, while the
amplitudes of the waves are reproduced quite accurately.

	In addition to a pointwise comparison between analytic and
numerical amplitudes, we can compute the energy loss through a large
sphere inclosing the black hole~\cite{cpm79,pp94}. For $\ell = 2$ this
is given by

\begin{equation}
\frac{dE}{dt} = \frac{1}{384\pi} 
\left( {\partial Q_{2} \over {\partial t}} \right)^2 \ . 
\end{equation}

\begin{figure}
\centerline{\epsfysize=12cm \epsfbox{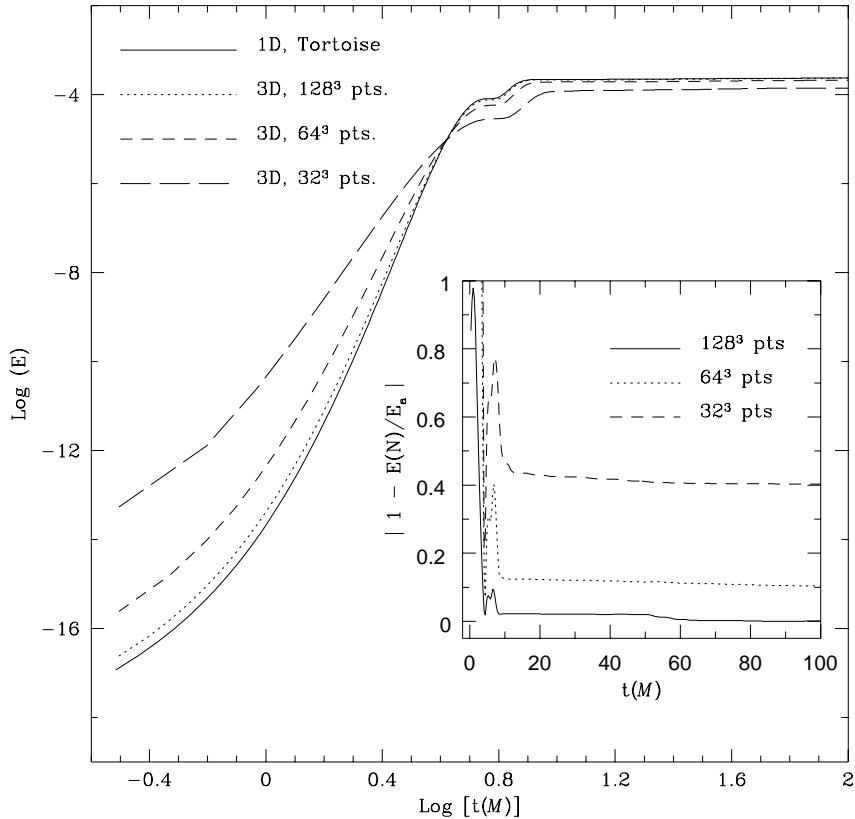}} 
\caption[fig4]{\label{fig4} 
Logarithm of the energy radiated through a spherical surface at
$R=15\;M$ for $\ell = 2$ as a function of time. The inset shows the 
absolute relative error between the expected radiated energy and the
computed one.
}
\end{figure}

	In Figure \ref{fig4} we show the total energy radiated through
a sphere of radius $R=15M$. Here, our numerical results converge to
the analytic one even for late times. This is because very little
energy is contained in the late-time oscillations  
which have significant phase errors and, later, amplitude errors.

	This becomes apparent in the inset of Figure \ref{fig4}, which
shows the absolute relative error between the expected radiated energy
and the computed one as a function of time and grid resolution. Apart
from the initial larger values related to the outgoing part of the
initial wave packet, the relative errors tend to reach constant values
of $40\%,\ 10\%$ and approximately $2\%$ for $(32)^3$, $(64)^3$ and
$(128)^3$ points respectively. From these estimates and from the fact
that we are measuring perturbations which have a typical wavelength
$\lambda \sim 17 M$, we can infer that a ratio $(h/\lambda)
\sim 10^{-2}$ on a uniform grid should be sufficient to provide a
physical description of gravitational wave propagation on a curved
background with an error smaller than a few percent. It should be
noted that relative errors larger than the ones discussed above for
the radiated energy can be measured when comparing the computed and
``analytic'' waveforms at a given time and location. However, these
errors appear only at later times, when the amplitude of the
perturbation is drastically reduced and its contribution to the total
radiated energy is negligible.

	Besides allowing us a somewhat more accurate simulation, for
any given resolution, the freezing boundary conditions
(\ref{freeze_bc}) have also solved the instability problems
encountered with the use of one-sided differencing at the inner
boundary. With ``freezing'' boundary conditions we were able to evolve
the code up to a physical time $t \approx 2 \times 10^5 M$ which
corresponds to $\approx 10^5$ crossing times. At that stage the $L_2$
norms of the relevant variables were showing a negative slope, clear
indication of stability of the code.

%We are confident that,
%with this prescription for the inner boundary conditions, this code
%would basically run ``for ever''.
%In Figure \ref{fig6} we show the timeseries of the logarithm
%of the $L_2$ norm of the Zerilli function $Q$ as computed with the 3D
%code and making use of $(32)^3$ gridpoints. The negative slope in the
%logarithmic scale for the late times of the timeseries (i.e. for $t
%\gtord 1000 M$) is a . It is
%interesting to notice from the timeseries of the the
%quasi-normal ringing of the black hole between $t \sim 10M$ and $t\sim
%100 M$, the following power-law tail of the late time behaviour of the
%perturbations \cite{gpp94,pp94} (despite the coarse resolution this is
%in good agreement with the theoretical value) and its interference
%with radiation reflected by the outer boundary. It is interesting to
%note that this occurs only for $t \gtord 500 M$ and thus allows for a
%significant dynamical range to study late time behaviour of
%perturbations in black hole spacetimes \cite{klp96}. We have stopped
%the computation after about one week of running time on 8 processors
%of the supercomputer Origin2000 at NCSA. At that stage the code had
%\begin{figure}
%\centerline{\epsfysize=12cm \epsfbox{norms_char.ps}} 
%\caption[fig6]{\label{fig6} 
%Timeseries of the logarithm of the $L_2$ norm of the Zerilli function
%$Q$ as computed with the 3D code and making use of
%characteristic boundary conditions at the inner boundary. The curve
%has been computed using $(32)^3$ gridpoints and almost one week of wall
%clock time.}
%\end{figure} 
	
	A final comment should be made about some modifications to the
present approach which could improve the quality of the numerical
simulations. As discussed in Sections \ref{tze} and \ref{nrsogr}, one
of the major difficulties encountered in this study are related to the
choice of a static slicing of the Schwarzschild background. A first
improvement can come from adopting a coordinate gauge that better
describes wave propagation in the vicinity of the horizon and that
avoids the ``freezing'' of the constant coordinate-time slices (see,
e.g.,~\cite{cs97b}). With a harmonic background, the wave
perturbations will propagate through the horizon in finite coordinate
time rather than piling up there.  Although the wave equation becomes
more complicated, it should be numerically straightforward to repeat
our test problem in this context.  A second independent improvement
could come from a more ingenious use of the memory resources. We have
shown that the use of high resolution grids improves the agreement
between the computed and the analytic solution. AMR techniques are
well known to be particularly suitable to study those configurations
in which a very high resolution is necessary only in some regions of
the computational domain. It is likely that their implementation in
the proximity of the horizon would provide the resolution necessary to
adequately resolve many of the details of the wave propagation.

%%%%%%%%%%%%%%%%%%%%%%%%%%%%%%%%%%%%%%%%%%%%%%%%%%%%%%%%%%%%%%%%%%%%%%
\section{Conclusion}
\label{conclusion}
%%%%%%%%%%%%%%%%%%%%%%%%%%%%%%%%%%%%%%%%%%%%%%%%%%%%%%%%%%%%%%%%%%%%%%

	We have presented 3D computations of gravitational radiation
scattering off a Schwarzschild black hole. This test problem
represents a useful tool to investigate numerical issues such as
finite differencing across the horizon, inner and outer boundary
conditions, evolution schemes, code parallelization, optimal use of
numerical resources and could be used as a standard benchmark on 3D
numerical relativity codes. The numerical results obtained from this
code are in good agreement with the ``analytical'' ones and converge
to the latter as the resolution is increased. Late time deviations
from the analytic solution are due to specifics of our choice of
coordinates.

	The present study has also allowed us to test the minimum
computational requirements of 3D numerical relativity computations
against the present resources available on modern parallel
computers. We have found that simple physical configurations such as a
perturbed Schwarzschild black hole can be handled quite satisfactorily
with most of the physical content of the solution being
reproduced. However, we also believe that much greater resources than
the ones available today may be necessary in order to study more
complex perturbative problems and, of course, the fully relativistic
evolution of binary black holes.

%%%%%%%%%%%%%%%%%%%%%%%%%%%%%%%%%%%%%%%%%%%%%%%%%%%%%%%%%%%%%%%%%%%%%%
\section*{Acknowledgments}
%%%%%%%%%%%%%%%%%%%%%%%%%%%%%%%%%%%%%%%%%%%%%%%%%%%%%%%%%%%%%%%%%%%%%%

This work was supported by the NSF Binary Black
Hole Grand Challenge Grant Nos. NSF PHY 93-18152/ASC 93-18152 (ARPA
supplemented).

%%%%%%%%%%%%%%%%%%%%%%%%%%%%%%%%%%%%%%%%%%%%%%%%%%%%%%%%%%%%%%%%%%%%%%
\appendix
%%%%%%%%%%%%%%%%%%%%%%%%%%%%%%%%%%%%%%%%%%%%%%%%%%%%%%%%%%%%%%%%%%%%%%

\section{Finite Difference Equations}
\label{afde}

	We here give explicit expressions for the finite difference
forms implemented in the second order accurate Macormack evolution
scheme.  We use a standard notation where lower indices refer to the
spatial location of the gridpoint and the upper index refers to the
time level. As mentioned in Section \ref{tze}, we solve for a set of 5
first order partial differential equations [i.e. (\ref{first_order})]
that have the symbolic form

\begin{equation}
{\partial A \over \partial t}  - \left (
{\partial B \over \partial x}  +
{\partial C \over \partial y}  +
{\partial D \over \partial z}  \right )
= {\rm RHS} (A,B,C) \ , 
\end{equation}
\begin{equation}
{\partial B \over \partial t}  - 
{\partial A \over \partial x} = 0 \ , \hskip 1.0truecm
{\partial C \over \partial t}  - 
{\partial A \over \partial y} = 0 \ , \hskip 1.0truecm
{\partial D \over \partial t}  - 
{\partial A \over \partial z} = 0 \ .
\end{equation}
All variables are first evolved from the time level $t=t_n$ to
the time level $t={\tilde t}_{n+1} = t_n + \Delta t$ by means of the
predictor step, in which the spatial derivatives are computed using a
first order accurate backward differencing:
\begin{eqnarray}
\label{a3}
\tilde{A}^{n+1}_{i,j,k} = A^n_{i,j,k} + \Delta t \Biggl\{
	\frac{1}{\Delta x}(B^n_{i,j,k} - B^n_{i-1,j,k}) + 
	\frac{1}{\Delta y}(C^n_{i,j,k} - C^n_{i,j-1,k}) + 
	\frac{1}{\Delta z}(D^n_{i,j,k} - D^n_{i,j,k-1}) + \\ \nonumber
	{\rm RHS} (A^n_{i,j,k},B^n_{i,j,k},C^n_{i,j,k})
	\Biggr\} \ ,
\end{eqnarray}
\begin{equation}
\tilde{B}^{n+1}_{i,j,k} = B^n_{i,j,k} + \frac{\Delta t}{\Delta x} 
	\left( A^n_{i,j,k} - A^n_{i-1,j,k} \right) \ ,
\end{equation}
\begin{equation}
\tilde{C}^{n+1}_{i,j,k} = C^n_{i,j,k} + \frac{\Delta t}{\Delta y}
	\left(A^n_{i,j,k} - A^n_{i,j-1,k} \right) \ ,
\end{equation}
\begin{equation}
\tilde{D}^{n+1}_{i,j,k} = D^n_{i,j,k} + \frac{\Delta t}{\Delta z}
	\left(A^n_{i,j,k} - A^n_{i,j,k-1}\right) \ .
\end{equation}

New, second order accurate values of the variables at the time level
$t=t_{n+1}$ are then computed with the corrector step, in which first
order forward differencing is used for the spatial derivatives:
\begin{eqnarray}
\label{a7}
A^{n+1}_{i,j,k} = \frac{1}{2}\Biggl\{
	A^n_{i,j,k} + \tilde{A}^{n+1}_{i,j,k} + \Delta t\Biggl[
	\frac{1}{\Delta x}(\tilde{B}^{n+1}_{i+1,j,k} - 
	\tilde{B}^{n+1}_{i,j,k}) +
	\frac{1}{\Delta y}(\tilde{C}^{n+1}_{i,j+1,k} - 
	\tilde{C}^{n+1}_{i,j,k}) + \\ \nonumber
	\frac{1}{\Delta z}(\tilde{D}^{n+1}_{i,j,k+1} - 
	\tilde{D}^{n+1}_{i,j,k}) + 
	{\rm RHS} (\tilde{A}^n_{i,j,k},\tilde{B}^n_{i,j,k},
	\tilde{C}^n_{i,j,k})
	\Biggr]\Biggr\} \ ,
\end{eqnarray}

\begin{equation}
B^{n+1}_{i,j,k} = \frac{1}{2}\left[
	B^n_{i,j,k} + \tilde{B}^{n+1}_{i,j,k} + 
	\frac{\Delta t}{\Delta x} \left( \tilde{A}^{n+1}_{i+1,j,k} 
	- \tilde{A}^{n+1}_{i,j,k} \right)\right] \ , 
\end{equation}
\begin{equation}
C^{n+1}_{i,j,k} = \frac{1}{2}\left[
	C^n_{i,j,k} + \tilde{C}^{n+1}_{i,j,k} + 
	\frac{\Delta t}{\Delta y} \left( \tilde{A}^{n+1}_{i,j+1,k} 
	- \tilde{A}^{n+1}_{i,j,k} \right)\right] \ , 
\end{equation}
\begin{equation}
D^{n+1}_{i,j,k} = \frac{1}{2}\left[
	D^n_{i,j,k} + \tilde{D}^{n+1}_{i,j,k} + 
	\frac{\Delta t}{\Delta z} \left( \tilde{A}^{n+1}_{i,j,k+1} 
	- \tilde{A}^{n+1}_{i,j,k} \right)\right] \ . 
\end{equation}
The first order error term introduced in the predictor step is exactly
cancelled in the corrector step, so that the algorithm becomes second
order accurate. The fifth equation of (\ref{first_order}) involves a
simple time integration; its finite difference form can be deduced
from (\ref{a3}) and (\ref{a7}) when all of the spatial derivatives are
taken to be zero.

\section{Outer Boundary Conditions}
\label{aobcs}

	We here present second order accurate finite difference
expressions for the outgoing-wave Sommerfeld conditions of the form
(\ref{somm_cart_rad}). In general, these conditions need to be applied
at the six external faces of the cubical grid, but here we will
concentrate only on the expressions for the $(x_{_{Max}},y,z)$ plane,
from which equivalent expressions for the other outer planes can be
derived.

	After some manipulations involving a correct centering in time
and space of the relevant variables, the outer boundary conditions
(\ref{somm_cart_rad}) for the gridpoints $(i_{_M},j,k)$ on the
$(x_{_{Max}},y,z)$ plane assume the form
\begin{eqnarray}
\Phi^{n+1}_{i_{_M},j,k} = 
\left({1 \over 1 + F + \Delta t/ \langle R \rangle}\right) 
\Biggl[
\Phi^{n}_{i_{_M}-1,j,k}  \left(1 + F -
{\Delta t \over {\langle R \rangle}} \right) +
\Phi^{n}_{i_{_M},j,k}    \left(1 - F - 
{\Delta t \over {\langle R \rangle}} \right) - \\ \nonumber
\Phi^{n+1}_{i_{_M}-1,j,k}\left(1 - F + 
{\Delta t \over {\langle R \rangle}} \right) 
\Biggr] \ , 
\end{eqnarray}
where 
\begin{equation}
F = {\langle R \rangle \over \langle x \rangle } {\Delta t \over h} 
\end{equation}
and 
\begin{equation}
\langle R \rangle = {R_{i_{_M},j,k} - R_{i_{_M}-1,j,k} \over 2} \ , 
\hskip 3.0truecm
\langle x \rangle = {x_{i_{_M},j,k} - x_{i_{_M}-1,j,k} \over 2} \ .
\end{equation}
	In the case in which ``octant'' symmetries are not used, the
equivalent expressions for the gridpoints $(i_m,j,k)$ on the
$(x_{min},y,z)$ plane will be obtained by changing $(i_{_M},j,k)
\rightarrow (i_m,j,k)$, $(i_{_M}-1,j,k) \rightarrow (i_m+1,j,k)$ and $F
\rightarrow - F $. ($i_{_M}$ and $i_m$ are here the first and last grid
points in the $x$-direction respectively.)

%%%%%%%%%%%%%%%%%%%%%%%%%%%%%%%%%%%%%%%%%%%%%%%%%%%%%%%%%%%%%%%%%%%%%%


\begin{references}
\bibitem{note1}{Hereafter, with 3D simulations, we will refer to
	computations using the standard $3+1$ splitting of
	spacetime.}
\bibitem{fn2}{The spurious minimum observed at $t \approx
		10 M$ is an ``edge-effect'' produced by the low
		resolution at which these data have been obtained. It 
		gradually disappears as higher resolution is used.}
\end{references}
\end{document}